\documentclass[pra,10pt,twocolumn,superscriptaddress,showpacs]{revtex4-1}

\usepackage{epsfig}
\usepackage{amsmath}
\usepackage{graphicx}
\usepackage{graphics}
\usepackage{float}
\usepackage{amssymb}
\usepackage[usenames,dvipsnames]{color}
\usepackage{natbib}
\usepackage{epstopdf}
\usepackage{verbatim}
\usepackage{wrapfig}
\usepackage{mathtools}

\graphicspath{{Figure/}}

\usepackage{bm}

\newcommand{\bra}[1]{\left\langle #1\right|}
\newcommand{\ket}[1]{\left| #1\right\rangle}

\newcommand{\opav}[3]{\langle #1 | #2 | #3 \rangle}

\newcommand{\beq}{\begin{equation}}
\newcommand{\eeq}{\end{equation}}
\newcommand{\tr}{\text{Tr}}

\begin{document}

\title{Effect of initial system-environment correlations with spin environments}

\author{Mehwish Majeed}
\affiliation{School of Science \& Engineering, Lahore University of Management Sciences (LUMS), Opposite Sector U, D.H.A, Lahore 54792, Pakistan}

\author{Adam Zaman Chaudhry}
\email{adam.zaman@lums.edu.pk}
\affiliation{School of Science \& Engineering, Lahore University of Management Sciences (LUMS), Opposite Sector U, D.H.A, Lahore 54792, Pakistan}

\begin{abstract}

Understanding the dynamics of open quantum systems is a highly important task for the implementation of emerging quantum technologies. To make the problem tractable theoretically, it is common to neglect initial system-environment correlations. However, this assumption is questionable in situations where the system is interacting strongly with the environment. In particular, the system state preparation can then influence the dynamics of the system via the system-environment correlations. To gain insight into the effect of these correlations, we solve an exactly solvable model of a quantum spin interacting with a spin environment both with and without initial correlations for arbitrary system-environment coupling strengths. We show that the effect of the system state preparation may or may not be significant in the strong system-environment coupling regime at low temperatures. We also study the dynamics of the entanglement between two spins interacting with a common spin environment with and without initial system-environment correlations to demonstrate that the correlations can play a significant role in the dynamics of two-qubit systems as well. 

\end{abstract}

\pacs{03.65.Yz, 05.30.-d, 03.67.Pp, 42.50.Dv}

\maketitle


\section{Introduction}  

Realistic quantum systems interact with their environment. This means that describing the dynamics of such open quantum systems is a highly non-trivial problem. Various techniques have been formulated for this task which generally employ a variety of approximations and assumptions in order to make the complicated dynamics computationally feasible \cite{BPbook,Weissbook}. For instance, it is generally assumed that the system and the environment are weakly interacting so that perturbation theory can be employed. The environment is often assumed to have a very short correlation time so that memory effects are negligible. Moreover, the system and the environment are assumed to be initially in a product state, with the environment in a thermal equilibrium state. In other words, initial system-environment correlations are completely neglected, the justification being that the system-environment correlations can be ignored if the system and environment are weakly interacting. For Markovian environments, we can expect that the environment quickly loses any information regarding the system, thereby providing further justification for ignoring the system-environment correlations \cite{Modi2011}. All these approximations are expected to break down in the strong system-environment coupling regime. 

With various quantum systems of practical interest such as superconducting qubits, quantum dots and light-harvesting complexes exhibiting strong interactions with the environment, various studies have been performed to critically analyze the effect of the system-environment correlations \cite{HakimPRA1985, HaakePRA1985, Grabert1988, SmithPRA1990, GrabertPRE1997, PazPRA1997, LutzPRA2003, BanerjeePRE2003, vanKampen2004, BanPRA2009, HanggiPRL2009, UchiyamaPRA2010, TanimuraPRL2010, SmirnePRA2010, DajkaPRA2010, ZhangPRA2010,TanPRA2011, CKLeePRE2012,MorozovPRA2012, SeminPRA2012, LainePRL2012, ChaudhryPRA2013a,ChaudhryPRA2013b,ChaudhryCJC2013,FanchiniSciRep2014,BuscemiPRL2014,FanSciRep2015,ChenPRA2016,VegaRMP2017,VegaPRA2017,ShibataJPhysA2017,CaoPRA2017}. Unfortunately, the effect of the initial correlations are only expected to be very significant in the strong system-environment coupling regime, which is not amenable to the usual perturbative methods \cite{BPbook}. To counter this problem, one approach has been to study the effect of the initial correlations using exactly solvable models - see, for example, Refs.~\cite{MorozovPRA2012} and \cite{ChaudhryPRA2013a}. However, these exactly solvable models in turn have different limitations. In particular, the study of initial correlations performed in Refs.~\cite{MorozovPRA2012} and \cite{ChaudhryPRA2013a} use exactly solvable dephasing models where the system energy does not change. In other words, the diagonal elements of the system density matrix remain unchanged. As another example, Ref.~\cite{SmirnePRA2010} studied the effect of initial correlations in the Jaynes-Cummings model, where the diagonal elements of the central qubit do change, but the `environment' is only a single harmonic oscillator. 

Our objective in this work is to examine an exactly solvable model in which we can include the effects of initial system-environment correlations exactly and in which both the diagonal and off-diagonal elements of the density matrix change. In other words, our system undergoes dephasing and its energy also changes due to its interaction with the environment. We believe that studying such a model will give useful insights into the role of the initial system-environment correlations just like previous works (see, for example, Refs.~\cite{SmirnePRA2010,MorozovPRA2012,ChaudhryPRA2013a}) have done before. To this end, we examine an extension of a previously studied model of a single spin interacting with an environment consisting of a collection of spins \cite{CucchiettiPRA2005}. The system spin Hamiltonian does not commute with the system-environment interaction, making the solution non-trivial. The system and the environment are allowed to reach a joint equilibrium state, and then a projective measurement is performed on the system to prepare the desired system state. The joint system-environment equilibrium state is, in general, a correlated system which is different from the usually assumed uncorrelated product state of the system and environment \cite{CKLeePRE2012}. The state preparation influences the subsequent dynamics of the system spin via the system-environment correlations that existed before the state preparation \cite{MorozovPRA2012,ChaudhryPRA2013a,ChaudhryPRA2013b,ShibataJPhysA2017}. The advantage of this model is that we obtain relatively simple expressions for the evolution of the Bloch vector of the system spin for arbitrary temperature and arbitrary system-environment coupling strength with both initially uncorrelated and correlated system-environment states. The exact analytical solutions for the Bloch vector allow us to show that the state preparation can have a very significant influence on the system dynamics via the system-environment correlations. As expected, we find that with relatively high temperatures and/or weak system-environment coupling strength, the effect of the initial correlations is negligible. For lower temperatures and stronger system-environment coupling strengths, the initial correlations can play a very significant role. However, interestingly, this is not always the case - even with very low temperatures and strong system-environment coupling strength, it is possible that the state preparation does not play any role in the system dynamics. This is in contrast with the harmonic oscillator environments investigated previously \cite{MorozovPRA2012, ChaudhryPRA2013a}. We then extend our model to two spins interacting with a common spin environment. We once again demonstrate that the initial correlations can play a very important role. In particular, the phenomena of entanglement sudden death and birth \cite{EberlyPRL2004,EberlyScience2007,EberlyScience2009,lopezPRL2008} can differ greatly due to the initial correlations.

This paper is organized as follows. In Sec.~II, we introduce our model of a spin interacting with a spin environment, and we solve the dynamics of the central spin with and without initial correlations. In Sec.~III, we extend our model to two central spins interacting with the spin environment, and find the dynamics of the entanglement between the two spins with and without initial correlations. Finally, we conclude in Sec.~IV.

\section{The model}

Our model consists of a single spin-1/2 particle interacting with a spin bath consisting of $N$ spin-1/2 particles. Our system-environment Hamiltonian is
\begin{equation}
H = H_{S} + H_{B} + H_{SB},
\end{equation}
 where $H_{S}$ and $H_{B}$, the self-Hamiltonians of the central system and the environment, are defined to be (we set $\hbar = 1$ throughout)
 \begin{equation}
 H_{S}=\dfrac{\varepsilon}{2}\sigma_{z} + \dfrac{\Delta}{2}\sigma_{x},
 \end{equation}
 and
  	\begin{equation}\label{2}
  	H_{B} = \sum_{i=1}^{N}\dfrac{\varepsilon_{i}}{2}\sigma_{z}^{(i)}+\sum_{{i}=1}^{N}\sigma_{z}^{(i)}\sigma_{z}^{(i+1)}\chi_{i},
  	\end{equation}
  	while the system-environment interaction is 
  	\begin{equation}\label{3}
  	H_{SB}=\dfrac{1}{2}\sigma_{z}\otimes\sum_{i=1}^{N}g_{i}\sigma_{z}^{(i)}.
  	\end{equation}
  	Here $\sigma_{k}$  $ (k = x, y, z) $ represent the usual Pauli spin matrices, and $\varepsilon$ and $\Delta$ denote the energy level spacing and the tunneling amplitude of the central two-level system respectively. Similarly, $\varepsilon_{i}$
  	denotes the energy level spacing for the $i^{\text{th}}$ environmental spin. We have also allowed the environment spins to interact each other via $\sum_{{i}=1}^{N} \sigma_{z}^{(i)}\sigma_{z}^{(i + 1)}\chi_{i}$, where $\chi_{i}$ characterizes the nearest-neighbor interaction strength between the environment spins. The central spin interacts with the environment spins through ${H}_{SB}$, where $g_i$ is the interaction strength between the central spin and the $i^{\text{th}}$ environment spin. Note that the system energy is not constant since $H_S$ does not commute with the total system-environment Hamiltonian.
  	
  	Our objective is to solve the dynamics of the central spin with both correlated and uncorrelated initial states. As such, we try to find the total system-environment unitary time-evolution operator. We first write the system-environment interaction Hamiltonian as $H_{\text{SB}} = \frac{1}{2}\sigma_z \otimes B$, where the environment operator $B$ is $B = \sum_{i = 1}^N g_i\sigma_z^{(i)}$. To proceed further, we follow a method similar to that in Ref.~\cite{CucchiettiPRA2005}. However, unlike Ref.~\cite{CucchiettiPRA2005}, we do not consider the environment Hamiltonian $H_B$ to be negligible and we will not in general assume the initial environment state to be a pure state. The eigenstates of $B$ can be written as products of the eigenstates $\ket{0_i}$ and $\ket{1_i}$  of the $i^{\text{th}}$ environment operator ${\sigma}_{z}^{(i)}$, where $\ket{0}$ denotes the spin `up' state and $\ket{1}$ the spin `down' state. As such, we write the eigenstates of $B$ as $\ket{n} \equiv \ket{n_1} \ket{n_2} \hdots \ket{n_N}$, with $n_i = 0, 1$. It is clear that 
\begin{equation}
{B}\ket{n} = G_{n}\ket{n},  
\end{equation}
with
\begin{equation}
G_{n}=\sum_{i=1}^{N}(-1)^{n_{i}}g_{i}.
\end{equation}
Also, since $B$ commutes with $H_B$, we expect that $H_B$ also has the same eigenstates. Indeed, 
\begin{equation}
\sum_{i = 1}^N \frac{\varepsilon_i}{2} \sigma_z^{(i)}\ket{n} = \frac{1}{2}\epsilon_n\ket{n}, 
\end{equation}
with 
\begin{equation}
\epsilon_{n} = \sum_{i=1}^{N}(-1)^{n_{i}}\varepsilon_{i},
\end{equation}
and 
\begin{equation}
\sum_{i = 1}^N \sum_{j = i + 1}^N \sigma_z^{(i)} \sigma_z^{(i+1)} \chi_{i} \ket{n} = \eta_{n}\ket{n}, 
\end{equation}
with
\begin{equation}
\eta_{n}=\sum_{i=1}^{N}(-1)^{n_{i}}(-1)^{n_{i+1}}\chi_{i}.
\end{equation}
Using the completeness relation over the states $\ket{n}$, that is, over all the different configurations of the environment spins, we find that the combined unitary time evolution operator for the system and the environment is
$$ U(t) = \sum_n e^{-i\epsilon_n t/2} e^{-i\eta_n t} e^{-i(H_S + H_{SB})t} \ket{n}. $$
This further simplifies since 
$$ e^{-i(H_S + H_{SB})t}\ket{n} = e^{-iH_{s,n}^{\text{eff}}t}\ket{n},$$
with $H_{s,n}^{\text{eff}} = \frac{\zeta_n}2\sigma_z + \frac{\Delta}{2}\sigma_x$, and $\zeta_n = \varepsilon + G_n$. We then have
\begin{equation}
\label{unitarytimeoperator}
	{U}(t)=\sum_{n=0}^{2^{N}-1}{U}_{n}(t)\ket{n}\bra{n},
\end{equation}
with
\begin{align*}
U_{n}(t)&={e^{-i\eta_{n}t}}{e^{-i\epsilon_{n}t/2}} \, \times \notag\\
&\left\lbrace\cos (\Omega_{n}t)-\dfrac{i}{\Omega_{n}}\sin(\Omega_{n}t)\left(\dfrac{\zeta_n}{2}\sigma_{z}+\dfrac{\Delta}{2}\sigma_{x}\right)\right\rbrace
\end{align*}
and $\Omega_{n}^{2}=\frac{1}{4}\left(\zeta_n^2 + \Delta^2\right)$. Eq.~\eqref{unitarytimeoperator} has the physical interpretation that for every configuration of environment spins $\ket{n}$, the effective dynamics of the central system can be found using $U_n(t)$.

To proceed further in finding the dynamics of the central system, we have to specify the initial system-environment state. The usual choice is to consider a simple product state of the form $\rho(0) = \rho_S(0) \otimes \rho_B$, with $\rho_S(0)$ the initial state of the system and $\rho_B$ the thermal state $e^{-\beta H_B}/Z_B$, with $Z_B = \tr_B [e^{-\beta H_B}]$ \cite{BPbook,Weissbook,pollakPRE2008}. However, this choice is not justified if the system and the environment are interacting strongly, since then the system-environment correlations can play a significant role. To take these correlations into account, we imagine that the system and the environment have been interacting strongly for a long time and have thus reached the joint equilibrium state proportional to $e^{-\beta H}$. A projective measurement is then performed on the system to prepare the desired initial system state $\ket{\psi}$. This then means that the initial system-environment state is now $\rho(0) = \ket{\psi}\bra{\psi}\otimes \opav{\psi}{e^{-\beta H}}{\psi}/Z$ \cite{MorozovPRA2012,ChaudhryPRA2013a,ChaudhryPRA2013b,ChaudhryCJC2013,ShibataJPhysA2017}. Although this is still a product state, the system-environment correlations that existed before the system state preparation have been taken into account. We now analyze the system dynamics with these two initial states one by one.

\subsection{\label{sec:level2A}Uncorrelated initial system-environment state}

The first choice of initial conditions is
\begin{equation}\label{UC1}
{{\rho}}(0)=\ket{\psi}\bra{\psi}\otimes\dfrac{{e^{-\beta{{H}_{B}}}}}{Z_{B}}.
\end{equation}  
We refer to this initial system-environment state as the `uncorrelated initial state' since the system-environment interaction before the state preparation is neglected. The system density matrix at time $t$ is $\rho_S(t) = \tr_B[e^{-iHt}\rho(0)e^{iHt}]$. Using Eq.~\eqref{unitarytimeoperator}, we find that the system density matrix with this initial state is 
\begin{equation}\label{URDM}
{{\rho}}_{S}(t)= \frac{1}{Z_B}\sum_{n=0}^{2^{N}-1} c_n U_{n}(t)\ket{\psi}\bra{\psi}U_n^\dagger,
\end{equation}
where $c_n = {e^{-\beta\eta_{n}}}{e}^{-{{\beta}\epsilon_{n}/2}}$, and $Z_{B}=\sum_n c_n$. This makes sense - each environment state configuration $\ket{n}$ occurs with probability $c_n/Z_B$ in the initial state, and for each configuration, the system dynamics is generated by $U_n(t)$. The total system state is then obtained simply by taking all the possible environment configurations into account. To quantify the dynamics of the central system, it is useful to find the Bloch vector components $p_k(t) = \tr_S[\sigma_k \rho_S(t)]$.  We find that the Bloch vector $\mathbf{p}(t)$ at time $t$ is given by $\mathbf{p}(t) = \frac{1}{Z_B}\mathbf{S}^{uc}(t)\mathbf{p}(0)$, that is, 
\begin{multline}
\left( {\begin{array}{c}
   p_x(t)\\
   p_y(t)\\
   p_z(t)
  \end{array} } \right) = \frac{1}{Z_B} \left( {\begin{array}{ccc}
   S^{uc}_{xx} & S^{uc}_{xy} & S^{uc}_{xz} \\
   S^{uc}_{yx} & S^{uc}_{yy} & S^{uc}_{yz} \\
   S^{uc}_{zx} & S^{uc}_{zy} & S^{uc}_{zz}
  \end{array} } \right) \left( {\begin{array}{c}
   p_x(0)\\
   p_y(0)\\
   p_z(0)
  \end{array} } \right)
  \end{multline}
with 
\begin{align}
S^{uc}_{xx}(t) &= \sum_{n} \frac{c_n}{4\Omega_n^2} \left[\zeta_n^2 \cos(2\Omega_n t) + \Delta^2 \right], \notag \\
S^{uc}_{xy}(t) &= -\sum_{n} \frac{c_n}{2\Omega_n} \zeta_n \sin(2\Omega_n t), \notag\\
S^{uc}_{xz}(t) &= \sum_{n} \frac{c_n}{2\Omega_n^2}\Delta \zeta_n \sin^2(\Omega_n t),\notag\\
S^{uc}_{yx}(t) &= \sum_{n} \frac{c_n}{2\Omega_n} \zeta_n \sin(2\Omega_n t), \notag\\
S^{uc}_{yy}(t) &= \sum_{n} c_n \cos(2\Omega_n t), \notag\\
S^{uc}_{yz}(t) &= -\sum_{n} \frac{c_n}{2\Omega_n} \Delta \sin(2\Omega_n t),\notag\\
S^{uc}_{zx}(t) &= \sum_{n} \frac{c_n}{2\Omega_n^2} \Delta \zeta_n \sin^2(\Omega_n t),\notag \\
S^{uc}_{zy}(t) &= \sum_{n} \frac{c_n}{2\Omega_n}\Delta \sin(2\Omega_n t),\notag \\
S^{uc}_{zz}(t) &= \sum_{n} \frac{c_n}{4\Omega_n^2} \left[\zeta_n^2 + \Delta^2 \cos(2\Omega_n t)\right].
\end{align}
Knowing the system-environment parameters, to find all the elements in this matrix one simply needs to perform sums over all the $2^N$ different environment configurations. We emphasize that this is an exact solution which is also valid if the $g_i$ are large. Furthermore, it is obvious that, in general, both the off-diagonal and diagonal elements of the system density matrix evolve.

\begin{figure}[t]
 			\includegraphics[scale = 0.6]{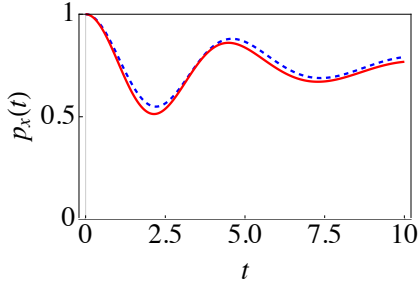}
 				\centering
				\caption{(Color online) Graph of $p_x(t)$ versus time $t$ for relatively weak system-environment coupling without initial correlations (dashed, blue line) and with initial correlations (solid, red line). We are working in dimensionless units with $\hbar = 1$ and we have set $\Delta = 1$. For simplicity, we have chosen the coupling strength $g_i$ and level spacing $\varepsilon_i$ to be the same for every environment spin. Here we have $g_i = 0.1$, $\varepsilon = 2$, $\varepsilon_i = 1$, $\beta = 1$, $\chi_i = 0$, and $N = 50$. The initial system state is specified by $p_x(0) = 1$.}
				 \label{weakcouplingillus}
			\end{figure}

				\begin{figure}[t]
				 \includegraphics[scale = 0.6]{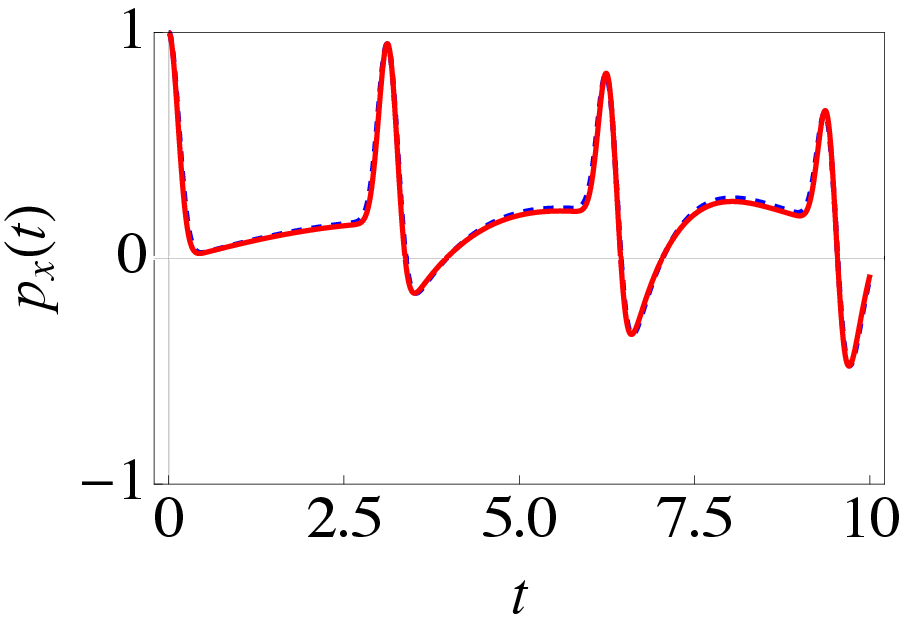}
 				\centering	
				 \caption{(Color online) Same as Fig.~\ref{weakcouplingillus}, except that now we have $\beta = 0.1$ and $g_i = 1$.}
				 \label{hightempillus}
			\end{figure}
			
			\begin{figure}[t]
				 \includegraphics[scale = 0.6]{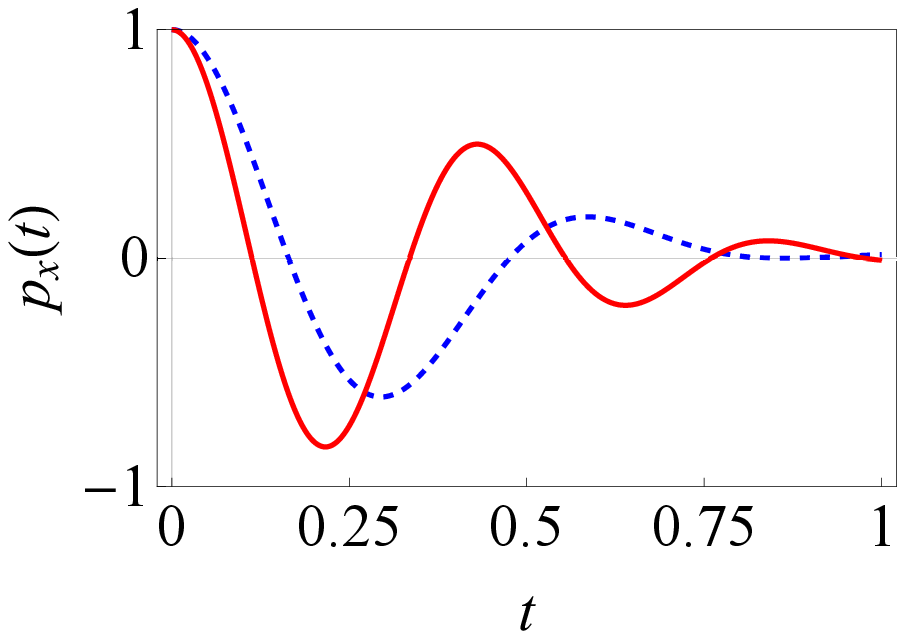}
 				\centering	
				 \caption{(Color online) Same as Fig.~\ref{weakcouplingillus}, except that now we have $\beta = 1$ and $g_i = 0.5$.}
				 \label{modtempmodc}
			\end{figure}

\subsection{\label{sec:level2A}Correlated initial system-environment state}

We now take the system-environment correlations (before the system state preparation) into account. The initial system-environment state is
\begin{equation}
\label{C1}
{\rho}(0)= \ket{\psi}\bra{\psi} \otimes \frac{\opav{\psi}{e^{-\beta{{H}}}}{\psi}}{Z},  
\end{equation}  
 where $Z=\tr_{S,B}[\ket{\psi}\bra{\psi}\otimes\opav{\psi}{e^{-\beta{H}}}{\psi}]$ is the partition function of the system and the environment as a whole.  We refer to this initial system-environment state as the `correlated initial state' since the system-environment interaction before the state preparation is taken into account. Note that the initial environment state ${\rho}_B=\opav{\psi}{e^{-\beta H }}{\psi}/Z$ depends on the ${H}_{SB}$ interaction term as well as the initial state preparation of the system and is thus not the canonical equilibrium state for the environment. It is obvious that if the system-environment coupling is small, the initial system-environment state (after the system state preparation) would be the same as that in Eq.~\eqref{UC1}. In other words, for weak system-environment coupling, the effect of the initial correlations is negligible. Furthermore, the state preparation influences the initial environment state due to the initial correlations, which means that the effect of the initial correlations also depends on the system state prepared. 
 
 To now work out the system dynamics, we start by observing that  
\begin{equation}
\sum_n e^{-\beta H}\ket{n}\bra{n} = \sum_n U_n(t = -i\beta)\ket{n}\bra{n},
\end{equation}
with $U_n(t)$ given in Eq.~\eqref{unitarytimeoperator}. This then allows us to write $Z = \sum_n c_n A_n$, with
\begin{align}
A_n = \cosh(\beta \Omega_n) - \frac{\sinh(\beta \Omega_n)}{\Omega_n}\opav{\psi}{\left(\frac{\zeta_n}{2}\sigma_z + \frac{\Delta}{2}\sigma_x \right)}{\psi}. \notag \\
\end{align}
The system density matrix at time $t$ is again given by $\rho_S(t) = \tr_B [e^{-iHt}\rho(0)e^{iHt}]$, but now with the initial state $\rho(0)$ given by Eq.~\eqref{C1}. Using Eq.~\eqref{unitarytimeoperator} for the time-evolution operator and simplifying, we find that the Bloch vector at time $t$ $\mathbf{p}(t)$ is now given by $\mathbf{p}(t) = \frac{1}{Z}\mathbf{S}^{c}(t)\mathbf{p}(0)$, with 
\begin{align}
S^{c}_{xx}(t) &= \sum_{n} \frac{c_n A_n}{4\Omega_n^2} \left[\zeta_n^2 \cos(2\Omega_n t) + \Delta^2 \right], \notag\\
S^{c}_{xy}(t) &= -\sum_{n} \frac{c_n A_n}{2\Omega_n} \zeta_n \sin(2\Omega_n t), \notag\\
S^{c}_{xz}(t) &= \sum_{n} \frac{c_n A_n}{2\Omega_n^2}\Delta \zeta_n \sin^2(\Omega_n t),\notag\\
S^{c}_{yx}(t) &= \sum_{n} \frac{c_n A_n}{2\Omega_n} \zeta_n \sin(2\Omega_n t), \notag\\
S^{c}_{yy}(t) &= \sum_{n} c_n A_n \cos(2\Omega_n t), \notag\\
S^{c}_{yz}(t) &= -\sum_{n} \frac{c_n A_n}{2\Omega_n} \Delta \sin(2\Omega_n t),\notag\\
S^{c}_{zx}(t) &= \sum_{n} \frac{c_n A_n}{2\Omega_n^2} \Delta \zeta_n \sin^2(\Omega_n t),\notag \\
S^{c}_{zy}(t) &= \sum_{n} \frac{c_n A_n}{2\Omega_n}\Delta \sin(2\Omega_n t), \notag\\
S^{c}_{zz}(t) &= \sum_{n} \frac{c_n A_n}{4\Omega_n^2} \left[\zeta_n^2 + \Delta^2 \cos(2\Omega_n t)\right].
\end{align}
Comparing with the uncorrelated case, we can see that the difference in the evolution is essentially because of the factor $A_n$ that takes into account the initial state preparation. Once again, this makes sense. The only difference compared to the usual uncorrelated case is due to the different initial environment state. Each environment spin configuration now occurs with probability $c_n A_n/Z$, as compared to $c_n/Z_B$ previously, and this is precisely what leads to the different Bloch vector evolution. We emphasize that our model allows us to find the effect of the correlations in an exact, non-perturbative manner, with both diagonals and off-diagonals of the two-level system changing. 

We now start to quantitatively analyze the difference in the evolution of the system state with and without initial correlations. Some general comments are in order. First, with weak system-environment coupling, as mentioned before, we expect that the evolution with the uncorrelated state and the correlated state will be very similar. Second, at high temperatures, we can again expect that the effect of the correlations is negligible - in the limit of very high temperatures, the system-environment state before the projective measurement on the system is a completely mixed state, meaning that there are no system-environment correlations. These two predictions are illustrated in Figs.~\ref{weakcouplingillus} and \ref{hightempillus} where we have plotted the Bloch vector component $p_x(t)$ starting from the uncorrelated and correlated system-environment state. For simplicity of presentation, we will be presenting the evolution of the Bloch vector component $p_x(t)$ only; however, all three components are generally changing. It is clear from the figures that there is a very small difference in the system dynamics due to the different initial states in these regimes.

		\begin{figure}[t]
				 \includegraphics[scale = 0.6]{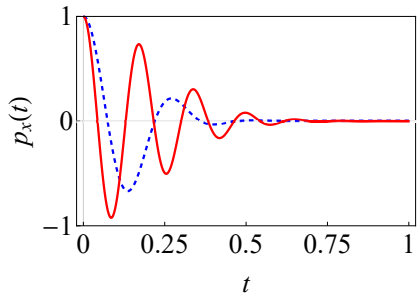}
 				\centering	
				 \caption{(Color online) Same as Fig.~\ref{weakcouplingillus}, except that now we have $\beta = 1$ and $g_i = 1$.}
				 \label{modtemphighc}
			\end{figure}

\begin{figure}[t]
 			\includegraphics[scale = 0.6]{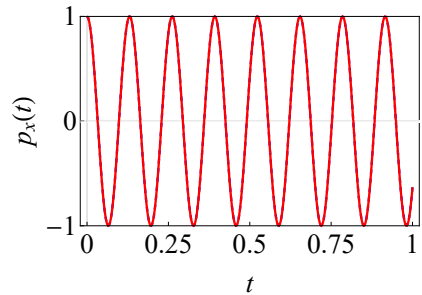}
 				\centering
				\caption{(Color online) Same as Fig.~\ref{weakcouplingillus}, except that now we have $g_i = 1$ and $\beta = 10$.}
				 \label{lowtemphighc}
			\end{figure}

\begin{figure}[t]
 			\includegraphics[scale = 0.6]{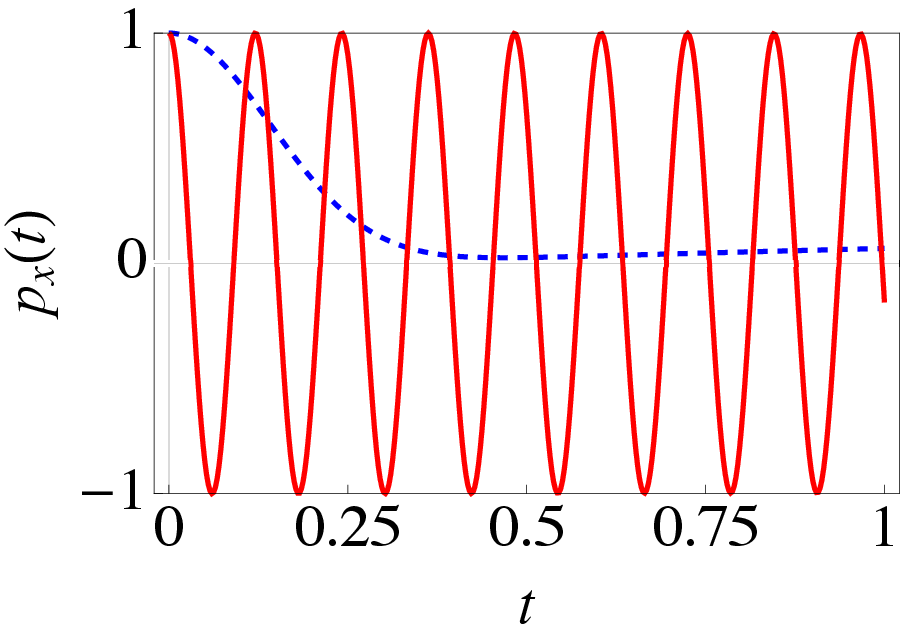}
 				\centering
				\caption{(Color online) Same as Fig.~\ref{weakcouplingillus}, except that now we have $\varepsilon_i = 0.01$, $g_i = 1$ and $\beta = 10$.}
				 \label{negHBlowtemp}
			\end{figure}

Let us consider now stronger coupling strengths. As expected, if the temperature is not high, with stronger system-environment strength, there is a more appreciable difference between the dynamics starting from the uncorrelated and correlated initial states. This is illustrated in Fig.~\ref{modtempmodc} 
where the coupling between the central spin and each of the environment spins has been set to $g_i = 0.5$ (the system parameters are $\varepsilon = 2$ and $\Delta = 1$). If the system-environment coupling strength is made even stronger, then there is an even bigger difference, as illustrated in Fig.~\ref{modtemphighc} where we have set $g_i = 1$. Proceeding along these lines, it is interesting to investigate what happens at even lower temperatures. Surprisingly, as illustrated in Fig.~\ref{lowtemphighc}, the difference in the dynamics due to state preparation disappears at lower temperatures, even for strong system-environment coupling strengths. This is contrary to the expectation that strong coupling strengths and low temperatures imply a greater effect of the initial system-environment correlations as is the case with harmonic oscillator environments \cite{MorozovPRA2012,ChaudhryPRA2013a}. However, the explanation is simple. Consider first the `uncorrelated' case $\rho(0) = \rho_S(0) \otimes e^{-\beta H_B}/Z_B$. At low temperatures, the environment will be (approximately) in its ground state. Considering all $\varepsilon_i$ to be positive, this means that the initial environment state will be $\ket{11\hdots 1}$, that is, all the environment spins will be in the spin `down' state. The system state, on the other hand, is simply the state that we choose to prepare $\ket{\psi}$. Now look at the initial state $\rho(0) = \rho_S(0) \otimes \opav{\psi}{e^{-\beta H}}{\psi}/Z$. At low temperatures, the system-environment state just before the system state preparation will be (approximately) the ground state of the total system-environment Hamiltonian $H$. If $H_B$ contributes significantly towards the total Hamiltonian, then the ground state corresponds to (approximately) the environment being all spins down and the system is spin up (assuming $g_i$ to be positive). Thus, the measurement on the system that prepares the initial system state does not affect the environment state, and the initial system-environment state is the same as before, meaning that the dynamics from the two initial states is the same. To test this prediction, let us instead consider the situation where $H_B$ is relatively small, which we can do by setting $\varepsilon_i$ to be small. Then, if the system-environment interaction Hamiltonian is significant, the ground state of the system-environment is not simply $\ket{0} \otimes \ket{1\hdots 1}$, that is, the system is in the spin `up' state and all the environment spins are spin `down'. Rather, the ground state is now a mixture of $\ket{0} \otimes \ket{1\hdots 1}$ and $\ket{1} \otimes \ket{0\hdots 0}$. On the other hand, the initial environment state with the uncorrelated system-environment state is the maximally mixed state. Clearly then, we expect a difference in the dynamics now. This is precisely what is illustrated in Fig.~\ref{negHBlowtemp}.

\begin{figure}[t]
	\includegraphics[scale = 0.6]{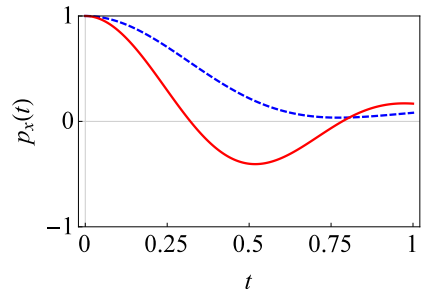}
	\centering	
	\caption{(Color online) Graph of $p_x(t)$ versus time $t$ for moderate system-environment coupling without initial correlations (dashed, blue line) and with initial correlations (solid, red line). Here we have considered the interactions between the spins of environment and we have set $\Delta = 1$. For simplicity, we have chosen the coupling strength $g_i$, level spacing $\varepsilon_i$ and interactions between the spins of environment $\chi_{i}$ to be the same for every environment spin. Here we have $g_i = 1$, $\varepsilon = 2$, $\varepsilon_i = 1$, $\beta = 1$, $N = 10$ and $\chi_{i}=0.1$. The initial system state is specified by $p_x(0) = 1$.}
	\label{10Beta1TLES1g1x01}
\end{figure}

\begin{figure}[t]
	\includegraphics[scale = 0.6]{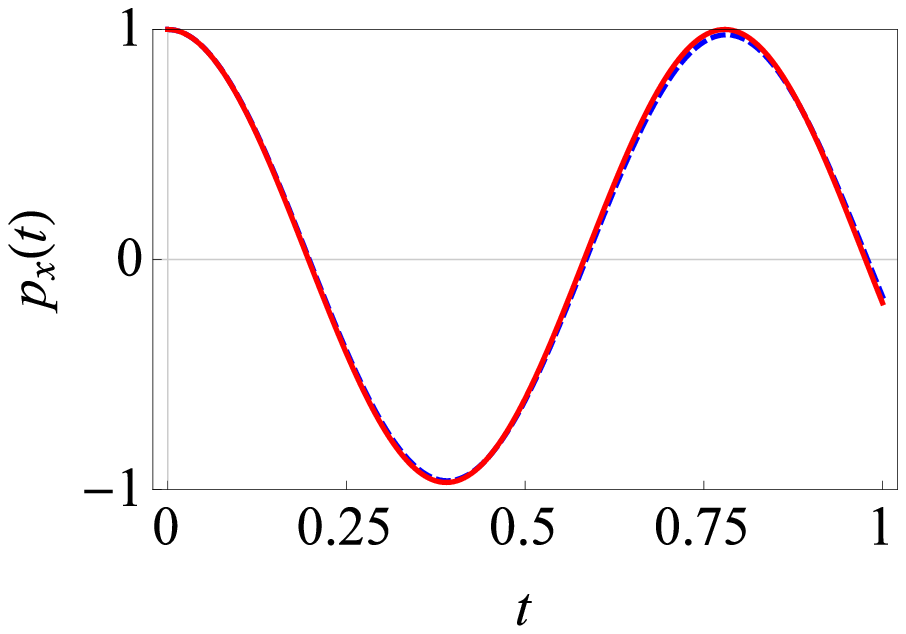}
	\centering	
	\caption{(Color online) Same as Fig.~\ref{10Beta1TLES1g1x01}, except that now we have $\beta = 10$ and $g_i=1$.}
	\label{10Beta10TLES1g1x01}
\end{figure}

\begin{figure}[t]
	\includegraphics[scale = 0.6]{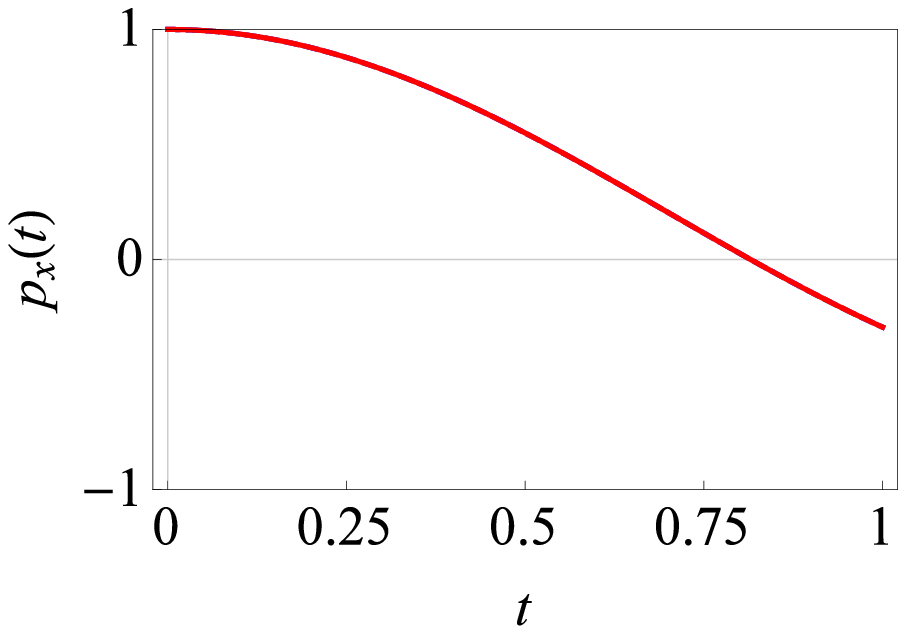}
	\centering	
	\caption{(Color online) Same as Fig.~\ref{10Beta1TLES1g1x01}, except that now we have $\beta = 10$, $\varepsilon_i=0.01$, $\chi_i=1$ and $g_i=1$.}
	\label{10Beta10TLES01g1x5}
\end{figure}

\begin{figure}[t]
	\includegraphics[scale = 0.6]{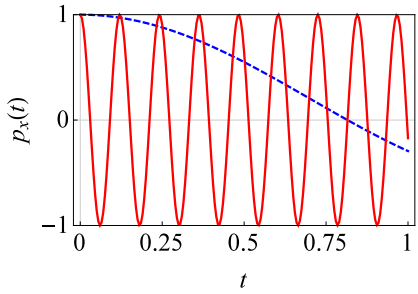}
	\centering	
	\caption{(Color online) Same as Fig.~\ref{10Beta1TLES1g1x01}, except that now we have $\beta = 10$, $\chi_i=1$ and $g_i=5$.}
	\label{positiveinteraction}
\end{figure}

		Next, we consider the environment spins to be interacting as well. As expected, for high temperatures and weak system-environment coupling, the initial correlations still have no effect on the system dynamics. On the other hand, as before, for strong system-environment coupling and moderate temperature, the initial correlations can play a significant role [see Fig.~\ref{10Beta1TLES1g1x01}]. Furthermore, as before, at lower temperatures, the difference in the system evolution with the uncorrelated and uncorrelated states can disappear as illustrated in Figs.~\ref{10Beta10TLES1g1x01} and \ref{10Beta10TLES01g1x5}. However, the situation is more complicated in Fig. \ref{positiveinteraction}, since now the initial correlations can play a role for very low temperatures. Let us try to explain this. Consider the spin-spin interaction for the environment $\chi_i$ to be positive, that is, the interaction is anti-ferromagnetic. Then, there are three effects at play here. First, due to the energies $\varepsilon_i$, the environment spins would like to be aligned. Second, if $g_i$ is positive, the environment spins would again like to be aligned. Third, due to the interaction between the spins, the environment spins would like to be anti-aligned. The different initial states can lead to different dynamics depending on which term is more dominant. If $\chi_i$ is small and $\varepsilon_i$ is relatively large, then combined system-environment ground state is approximately $\ket{0} \otimes \ket{1\hdots 1}$, that is, all the environment spins are aligned, which means that there is no difference in dynamics [see Fig.~\ref{10Beta10TLES1g1x01}]. On the other hand, if $\varepsilon_i$ is small and $\chi_i$ is large, the environment state consists of anti-aligned spins in the uncorrelated case. If the system-environment coupling is not completely dominant, then the environment state is the same for the correlated case. Once again, there is no difference in the dynamics [see Fig.~\ref{10Beta10TLES01g1x5}]. Now consider the situation where $\varepsilon_i$ and $\chi_i$ are comparable, while the system-environment coupling strength $g_i$ is dominant. Then for the uncorrelated state, the environment state is `confused' between being aligned or anti-aligned. However, for the correlated initial state, the environment state consists of all spins aligned.  Clearly then, the system dynamics will be different as illustrated in Fig.~\ref{positiveinteraction}.   	
		
		\begin{figure}[h!]
			\includegraphics[scale = 0.6]{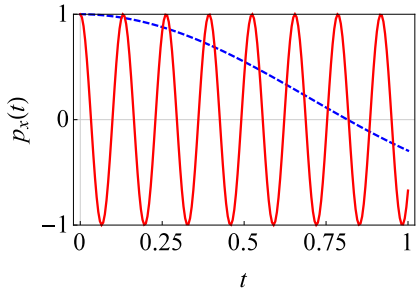}
			\centering
			\caption{(Color online) Graph of $p_x(t)$ versus time $t$ for relatively strong system-environment coupling without initial correlations (dashed, blue line) and with initial correlations (solid, red line). We are working in dimensionless units with $\hbar = 1$ and we have set $\Delta = 1$. The coupling strength $g_i$, level spacing $\varepsilon_i$  and interactions between the environment spins are considered to be Gaussian random variables. Here we have mean value of coupling strength $g_i = 5$ (standard deviation$=0.01$), $\varepsilon = 2$, mean value of level spacing $\varepsilon_i = 1$ (standard deviation$=0.001$), $\beta =10$, mean interactions between the spins $\chi_{i}=1$ (standard deviation$=0.01$) and $N = 10$. The initial state is specified by $p_x(0) = 1$.}
			\label{R10Beta10TLES1g5x1}
		\end{figure}

\begin{figure}[h!]
			\includegraphics[scale = 0.6]{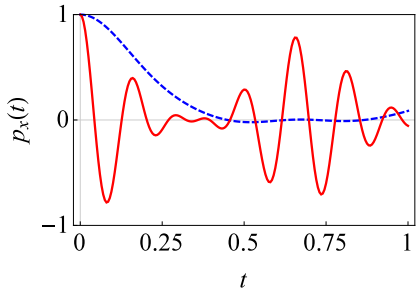}
			\centering
			\caption{(Color online) Same as Fig.~\ref{R10Beta10TLES1g5x1}, except that now the mean value of $g_i = 5$ (standard deviation$=1$), mean value of level spacing $\varepsilon_i = 1$ (standard deviation$=0.2$), $\beta =10$, and mean interactions between the spins $\chi_{i}=1$ (standard deviation$=0.2$).}
			\label{randomnoise}
		\end{figure}

		Until now, the numerical results we have presented have assumed that, for instance, the coupling strength between the central spin and each environment spin $g_i$ is the same. Of course, in reality this is unlikely to be the case. To overcome this shortcoming, we now illustrate that even if the environment parameters and the central spin-environment spin coupling strengths are randomly distributed, we obtain similar conclusions compared to what we have presented before. In Fig.~\ref{R10Beta10TLES1g5x1}, we have assumed that the system-environment coupling strength $g_i$ is a Gaussian random variable with a small standard deviation; the environment level spacings $\varepsilon_i$ and the inter-spin interactions $\chi_i$ are treated in a similar manner. We find that the difference between the initially correlated case and the uncorrelated case persists. This difference persists even with larger standard deviations [see Fig.~\ref{randomnoise}]. 
		
\section{Extension to two two level systems}
To further illustrate the difference in the system evolution with and without initial correlations, let us extend our formalism to deal with two qubits interacting with a common spin environment. It is well known that the dynamics of two qubits can display characteristics that are absent from the single qubit case. In particular, we can look at the behavior of the entanglement between the two qubits. If the initial state of the two qubits is a fully entangled state, then we know that due to the interaction with the environment, this entanglement can disappear in a finite amount of time, a phenomenon known as entanglement sudden death (ESD) \cite{EberlyPRL2004,EberlyScience2007,EberlyScience2009}. However, the entanglement between the two qubits can also revive - this is referred to as entanglement sudden birth (ESB) \cite{lopezPRL2008}. If the two qubits are not entangled to begin with, then we can investigate the dynamics of the generation of entanglement. 

Our Hamiltonian is a straightforward extension of the previous Hamiltonian for a single qubit. Namely, we now have 
\begin{equation}
 H = H_S^{(1)} + H_S^{(2)} + H_{12} + H_B + H_{SB}^{(1)} + H_{SB}^{(2)}, 
\end{equation}
with 
 \begin{equation}
 H_S^{(1)} = \frac{\varepsilon_1}{2}\sigma_{z1} + \frac{\Delta_1}{2}\sigma_{x1}, 
 \end{equation}
\begin{equation}
 H_S^{(2)} = \frac{\varepsilon_2}{2}\sigma_{z2} + \frac{\Delta_2}{2}\sigma_{x2},
\end{equation} 
\begin{equation}
H_{12} = \lambda \sigma_{z1}\sigma_{z2},
\end{equation} 
\begin{equation}
H_{SB}^{(1)}=\frac{1}{2}\sigma_{z1}\otimes\sum_{i=1}^{N}g_{i}\sigma_{z}^{(i)},
\end{equation}
\begin{equation}
H_{SB}^{(2)}=\frac{1}{2}\sigma_{z2}\otimes\sum_{i=1}^{N}g_{i}\sigma_{z}^{(i)},
\end{equation} \begin{equation}
H_{B}=\sum_{i=1}^{N}\dfrac{\varepsilon_{i}}{2}\sigma_{z}^{(i)}+\sum_{{i}=1}^{N}\sigma_{z}^{(i)}\sigma_{z}^{(i+1)}\chi_{i}.
\end{equation}
The qubits are labeled as $1$ and $2$ with two-level energies $\varepsilon_{1}$ and $\varepsilon_{2}$ and tunneling amplitudes $\Delta_{1}$ and $\Delta_{2}$ respectively, and are coupled by the interaction term $H_{12}$. $\sigma_{j1}$ and $\sigma_{j2}$ [with $(j=x,y,z)$] are the Pauli spin matrices for qubit $1$ and $2$ respectively. Our goal is to study the dynamics of the entanglement between the two qubits, starting from the uncorrelated and correlated initial states. To quantify the entanglement, we use the concurrence, defined as $C(t ) =$ max $(0,
\sqrt{M_{1}-M_{2}-M_{3}-M_{4}})$,  where $M_{i}$ are the eigenvalues of the matrix $ M = \rho_{S}(t)({\sigma}_{y1}\bigotimes\sigma_{y2})\rho^{*}_{S}(t)({\sigma}_{y1}\bigotimes{\sigma}_{y2})$, with $\rho^{*}_{S}(t)$ designating the complex conjugate of reduced density matrix of the two qubit system $\rho_{S}(t)$ \cite{WoottersPRL1998}. This quantity attains the maximum value of one for maximally entangled states and completely vanishes
for separable states.

\begin{figure}[t]
	\includegraphics[scale = 0.6]{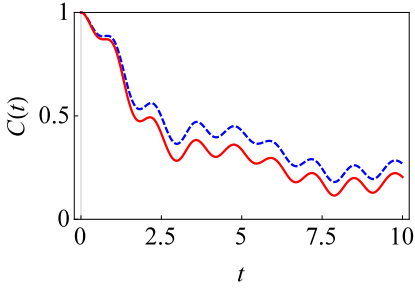}
	\centering
	\caption{(Color online) Graph of decay of entanglement of two qubits $C(t)$ versus time $t$ for relatively weak system-environment coupling without initial correlations (dashed, blue line) and with initial correlations (solid, red line). We are working in dimensionless units with $\hbar = 1$, and we have $\lambda=0$ and $\chi_i = 0$. For simplicity, we have chosen the coupling strength $g_i$ and  level spacing $\varepsilon_i$ to be the same for every environment. Here we have  $g_i = 0.1$, $\varepsilon_1 = 1$, $\varepsilon_2 = 2$, $\varepsilon_i = 1$, $\beta =1$, $\Delta_1 = 4$, $\Delta_2 = 1$ and $N = 50$. The initial state of two qubits is the maximally correlated state $\ket{\psi}=\frac{1}{\sqrt{2}}\left(\ket{0_1 0_2} + \ket{1_1 1_2}\right)$.}
	\label{D50CBeta1TLES1g01}
\end{figure}

\begin{figure}[t]
	\includegraphics[scale = 0.6]{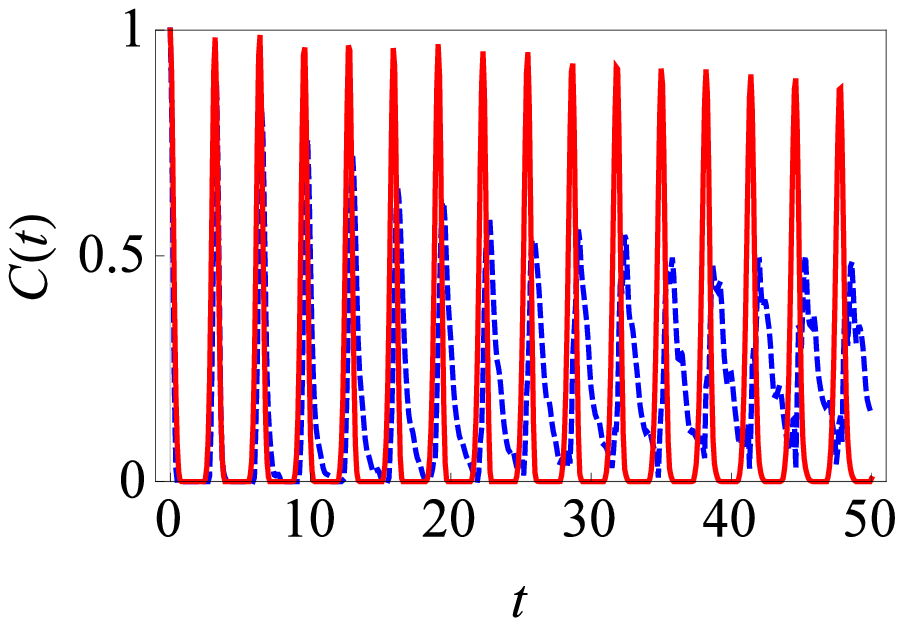}
	\centering
	\caption{(Color online) Same as Fig. \ref{D50CBeta1TLES1g01}, expect that now we have $\beta =1$ and $g_i = 0.5 $.}
	\label{D50CBeta1TLES1g05}
\end{figure}

 \begin{figure}[t]
 	\includegraphics[scale = 0.6]{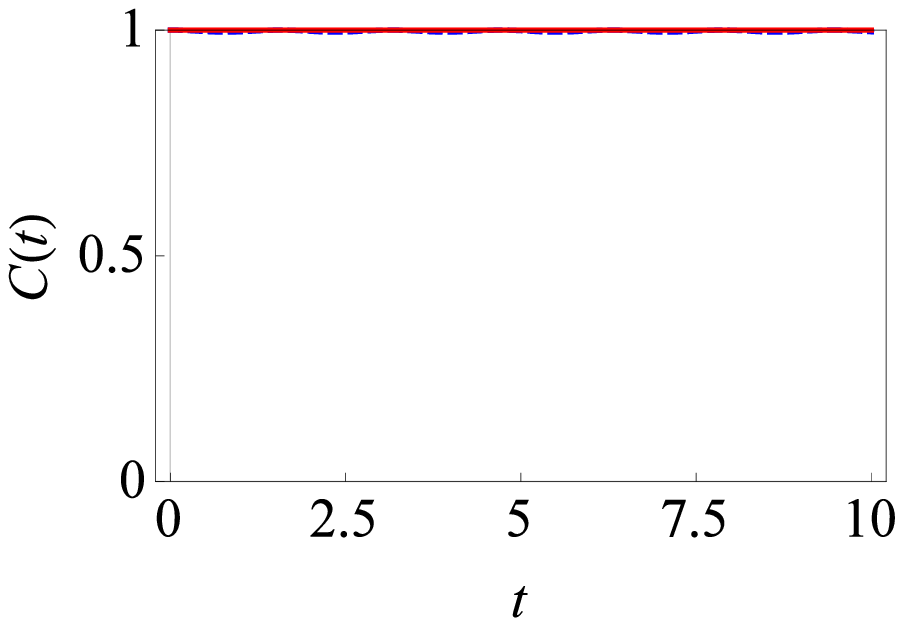}
 	\centering
 	\caption{(Color online) Same as Fig. \ref{D50CBeta1TLES1g01}, expect that now we have $\beta =10$ and $g_i = 1 $.}
 	\label{D50CBeta10TLES1g1}
 \end{figure}

  \begin{figure}[t]
 	\includegraphics[scale = 0.6]{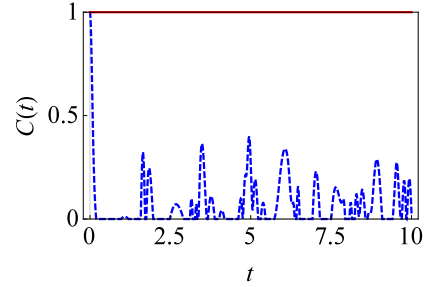}
 	\centering
 	\caption{(Color online) Same as Fig. \ref{D50CBeta1TLES1g01}, expect that now we have $\beta =10$, $\varepsilon_i=0.01$ and $g_i = 1 $.}
 	\label{D50CBeta10TLES01g1}
 \end{figure}
 
\begin{figure}[t]
	\includegraphics[scale = 0.6]{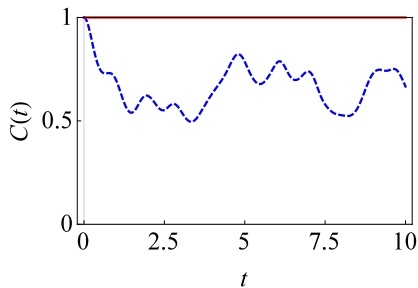}
	\centering
	\caption{(Color online) Same as Fig.~\ref{D50CBeta1TLES1g01}, except that now we have $\varepsilon_i = 0.01$, $\chi_i = 0.1$, $g_i = 1$, $\beta = 10$ and $N= 10$.}
	\label{D10CBeta10TLES01g1x01}
\end{figure}

We now find the two-qubit system density matrix starting from the uncorrelated state and from the correlated state. To simplify the presentation, let us first deal with the case $\lambda = 0$, that is, the two qubits are not directly interacting. To find the unitary time-evolution operator, we use a similar approach as before - we insert a completeness relation over the the eigenstates of the $H_B$. The total time-evolution operator can be written as 
\begin{align*} 
U(t) = &\sum_n e^{-i\epsilon_n t/2} e^{-i\eta_n t} \times \notag\\
&e^{-i(H^{(1)}_S + H^{(1)}_{SB})t}e^{-i(H^{(2)}_S + H^{(2)}_{SB})t} \ket{n}\bra{n},
\end{align*} 
with $\epsilon_n$ and $\eta_n$ as defined before. Now, 
\begin{equation*}
e^{-i(H^{(1)}_S + H^{(1)}_{SB})t}e^{-i(H^{(2)}_S + H^{(2)}_{SB})t}\ket{n} = e^{-iH_{s,n}^{(1)\text{eff}}t}e^{-iH_{s,n}^{\text{(2)eff}}t}\ket{n},
\end{equation*}
with $H_{s,n}^{(i)\text{eff}} = \frac{\zeta_{ni}}2\sigma_{zi} + \frac{\Delta_{i}}{2}\sigma_{xi}$, and $\zeta_{ni} = \varepsilon_{i} + G_n$ (${i=1,2}$). We then obtain 
\begin{equation}
\label{totalunitarytimeoperator}
	{U}(t)=\sum_{n=0}^{2^{N}-1}{U}_{n}^{(1)}(t){U}_{n}^{(2)}(t)\ket{n}\bra{n},
\end{equation}
where
\begin{align*}
U^{(i)}_{n}(t)&={e^{-i\eta_{n}t/2}}{e^{-i\epsilon_{n}t/4}} \, \times \notag\\
&\left\lbrace\cos (\Omega_{ni}t)-\dfrac{i}{\Omega_{ni}}\sin(\Omega_{ni}t)\left(\dfrac{\zeta_{ni}}{2}{\sigma}_{zi}+\dfrac{\Delta_i}{2}{\sigma}_{xi}\right)\right\rbrace,
\end{align*}
and $\Omega_{ni}^2=\frac{1}{4}\left(\zeta_{ni}^2 + \Delta_i^2\right)$. For the uncorrelated initial system-environment state [see Eq. (\ref{UC1})], it follows that 
\begin{align}\label{UCRDM}
\rho_{S}(t)&=\sum_{n=0}^{2^{N}-1} \frac{c_n}{Z_{B}}U^{(1)}_{n}(t)U^{(2)}_{n}(t)\ket{\psi}\bra{\psi}U_n^{(2)\dagger}U_n^{(1)\dagger},
\end{align}
with $Z_{B}=\sum_n c_n$. On the other hand, for the correlated initial system-environment state [see Eq.~\eqref{C1}], we obtain 
\begin{align}\label{CRDM}
{\rho}_{S}(t)&=\sum_{n=0}^{2^{N}-1} \frac{c_n A_n}{Z}U^{(1)}_{n}(t)U^{(2)}_{n}(t)\ket{\psi}\bra{\psi}U_n^{(2)\dagger}U_n^{(1)\dagger},
\end{align}
where $Z=\sum_n A_{n} c_n$, and $A_{n}=\langle \psi| A^{(1)}_{n}A^{(2)}_{n}|\psi\rangle $, with 
\begin{align}
A^{(i)}_n = \cosh(\beta \Omega_{ni}) - \frac{\sinh(\beta \Omega_{ni})}{\Omega_{ni}}\left[\frac{\zeta_{ni}}{2}\sigma_{zi} + \frac{\Delta_{i}}{2}\sigma_{xi}\right], \notag \\
\end{align}
appearing due to the initial correlations.

 \begin{figure}[t]
 	\includegraphics[scale = 0.6]{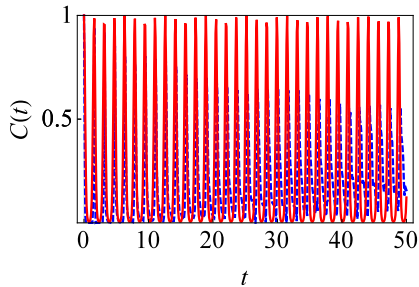}
 	\centering
 	\caption{(Color online) Same as Fig. \ref{D50CBeta1TLES1g01}, expect that now we have considered the interaction term $\lambda$ between the two qubits of system with $\lambda=3$, $\beta =1$ and $g_i = 1 $.}
 	\label{D50CBeta1TLES1g1L3}
 \end{figure}

\begin{figure}[t]
	\includegraphics[scale = 0.6]{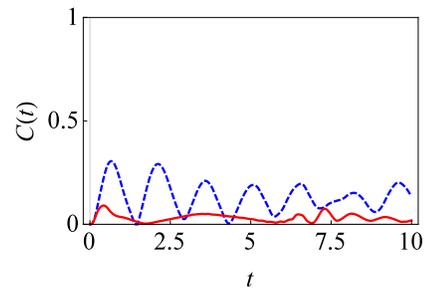}
	\centering
	\caption{(Color online) Graph of birth of entanglement of two qubits $C(t)$ versus time $t$ for relatively moderate system-environment coupling without initial correlations (dashed, blue line) and with initial correlations (solid, red line). We are working in dimensionless units with $\hbar = 1$ and we have set $\lambda=5$. Here we have  $g_i = 0.5$, $\varepsilon_1 = 1$, $\varepsilon_2 = 2$, $\varepsilon_i = 1$, $\beta =1$, $\Delta_1 = 4$, $\Delta_2 = 1$ and $N = 50$. The initial state of the two qubits is the product state $\ket{\psi}=\ket{0_1 0_2}$.}
	\label{D50CBBeta1TLES1g05L5}
\end{figure}

With the two-qubit density matrix in hand, we can look at the behavior of entanglement with and without initial correlations and show that there can be considerable differences. Let us first look at the weak system-environment coupling scenario. In this case, as expected, an initially entangled state largely loses its entanglement due to the interaction with the spin environment. As shown in Fig.~\ref{D50CBeta1TLES1g01}, there is a small difference between the correlated and uncorrelated cases. However, with stronger system-environment coupling, there can be a more significant difference with and without initial correlations in the entanglement dynamics [see Fig.~\ref{D50CBeta1TLES1g05}]. However, once again if we reduce the temperature further, the difference in the dynamics can disappear as shown in Fig.~\ref{D50CBeta10TLES1g1}. In fact, the concurrence is seen to remain very close to one. It is easy to explain why - the initial state of the environment with both choices of the initial state is the same, namely, all spins down, and the system-environment state remains (approximately) a product state. However, just like before, reducing the contribution of the environment Hamiltonian by decreasing the value of $\varepsilon_i$ can restore this difference [see Fig.~\ref{D50CBeta10TLES01g1}]. In fact, in this case, the uncorrelated initial state leads to repeated entanglement sudden death and birth, while with the correlated initial state, the entanglement remains largely intact. To further study the dynamics of central system, we can also consider interactions between the spins of the environment. For the uncorrelated initial state, the initial state of the environment at low temperatures will be all spins anti-aligned (assuming $\chi_i$ to be positive). For the correlated initial state, if the system-environment coupling is dominant, then we expect that the environment state would be all spins aligned. Thus, we expect a difference in the entanglement dynamics in this regime. This is precisely what is illustrated in Fig.~\ref{D10CBeta10TLES01g1x01}.

 We now investigate the entanglement dynamics if the qubits are directly interacting with each other, that is, $\lambda$ can now be non-zero. In this case, we have 
\begin{equation}
\label{lambdatotalunitarytimeoperator}
{U}(t)=\sum_{n=0}^{2^{N}-1}{U}_{n}^{(12)}(t)\ket{n}\bra{n},
\end{equation}
with
$${U}_{n}^{(12)}(t)={e^{-i\eta_{n}t}}{e^{-i\epsilon_{n}t/2}}e^{-i(H_{s,n}^{(1)\text{eff}}+H_{s,n}^{\text{(2)eff}}+H_{12})t}.$$
For the uncorrelated initial state [see Eq.(\ref{UC1})], the density matrix of system at some later time $t$ is
\begin{align}
{\rho}_{S}(t)=\frac{1}{Z_B}\sum_{n=0}^{2^{N}-1} c_n U^{(12)}_{n}(t)\ket{\psi}\bra{\psi}U_n^{(12)\dagger}(t),
\end{align}
 with $ Z_B=\sum_n c_{n}$. For the initially correlated state, we get 
\begin{multline}
{\rho}_{S}(t)=\frac{1}{Z}\sum_{n=0}^{2^{N}-1}c_n A_n {U^{(12)}_{n}(t)}\ket{\psi}\bra{\psi}{U^{(12)}_{n}}^{\dagger}(t)
\end{multline}
with
$ A_n = \opav{\psi}{e^{-\beta{(H_{s,n}^{(1)\text{eff}}+H_{s,n}^{\text{(2)eff}}+H_{12})}}}{\psi}$ and $Z=\sum_{n=0}^{2^{N}-1}c_{n}A_n$. For each $n$, we can calculate the $4 \times 4$ matrix $U_n^{(12)}$ numerically, and hence eventually the system density matrix. Once again, we can look at the entanglement dynamics starting from the uncorrelated and correlated system-environment states. Our central result - that there can be very significant differences between the dynamics due to the initial correlations - remains unchanged due to the presence of the qubit-qubit interaction [see Fig.~\ref{D50CBeta1TLES1g1L3}]. For completeness, we also illustrate this difference in Fig.~\ref{D50CBBeta1TLES1g05L5} for the case where the initial system state is a product state. In this case, it is is clear that the generation of entanglement is also impacted by the presence of the initial system-environment correlations.

\section{Conclusion}

 To conclude, we have solved the dynamics of a central two-level system interacting with a spin environment with and without initial system-environment correlations. For our model, both the diagonal and off-diagonal elements of the central spin density matrix evolve. We have found that as long as one remains in the high temperature and weak coupling regime, one can ignore any effect of initial correlations and system state preparation. On the other hand, for low temperatures and strong system-environment coupling strengths, the dynamics obtained from the correlated initial state and the uncorrelated initial state can be very different. However, surprisingly, this need not always be the case. We then extended our results to two spins interacting with a common spin environment to show that the entanglement dynamics can be affected by the initial correlations as well. Our results should lend very useful insights into the role of initial system-environment correlations.

\section*{Acknowledgments}
A.~Z.~C. acknowledges support from the LUMS FIF Grant FIF-413 and support from HEC under grant No 5917/Punjab/NRPU/R\&D/HEC/2016. 

\section*{ORCID iDs}
Adam Zaman Chaudhry https://orcid.org/0000-0002-4692-5792

\end{document}